\documentclass[journal]{IEEEtran}
\usepackage[pdftex]{graphicx}
\usepackage[T1]{fontenc}
\usepackage{lmodern}
\usepackage{textcomp}
\usepackage{latexsym}
\usepackage{cite}
\usepackage{bm}
\usepackage{amsmath}
\usepackage{amsfonts}
\usepackage[subrefformat=parens]{subcaption}
\usepackage{multirow}
\usepackage{booktabs}
\usepackage{array}

\newcommand{\vct}[1]{\mbox{\boldmath{$#1$}}}

\newcommand{\ee}        {\mathrm{e}}
\newcommand{\jj}        {\mathrm{j}}
\newcommand{\dd}        {\mathrm{d}}
\newcommand{\TT}        {\mathrm{T}}
\newcommand{\HH}        {\mathrm{H}}

\begin{document}
\title{Radar-Based Noncontact Measurement of Heartbeat of Humans and Chimpanzees Using Millimeter-Wave Radar with Topology Method}

\author{Itsuki~Iwata,~\IEEEmembership{Student Member, IEEE}, Takuya~Sakamoto,~\IEEEmembership{Senior Member, IEEE}, Takuya~Matsumoto, and~Satoshi~Hirata
  \thanks{I.~Iwata, and T.~Sakamoto are with the Department of Electrical Engineering, Graduate School of Engineering, Kyoto University, Kyoto 615-8510, Japan.}
  \thanks{T.~Matsumoto is with Faculty of Science, Shinshu University, Nagano 390-8621, Japan}
  \thanks{S.~Hirata is with Wildlife Research Center, Kyoto University, Kyoto 606-8203, Japan}}
\markboth{}%
{Iwata \emph{et al.}: Radar-Based Noncontact Measurement of Heartbeat of Humans and Chimpanzees Using Millimeter-Wave Radar with Topology Method}

\maketitle
\begin{abstract}
  This study proposes a method to determine the filter parameters required for the topology method, which is a radar-based noncontact method for measurement of heart inter-beat intervals. The effectiveness of the proposed method is evaluated by performing radar measurements involving both human participants and chimpanzee subjects. The proposed method is designed to enable setting of the filter cutoff frequency to eliminate respiratory components while maintaining the higher harmonics of the heartbeat components. Measurements using a millimeter-wave radar system and a reference contact-type electrocardiogram sensor demonstrate that the smallest errors that occur when measuring heart inter-beat intervals using the proposed method can be as small as 4.43 and 2.55 ms for humans and chimpanzees, respectively. These results indicate the possibility of using noncontact physiological measurements to monitor both humans and chimpanzees.
\end{abstract}


\IEEEpeerreviewmaketitle

\section{Introduction}
\IEEEPARstart{R}{adar} has been used to detect the body displacements caused by physiological activities such as respiration and heartbeats, thus allowing the physical and mental health conditions of humans and animals to be monitored in a noncontact manner. Noncontact monitoring of animals is especially important when the captive animal has a health condition but cannot communicate with its human caretakers. In addition, use of these technologies can contribute to the well-being of animals because they avoid the stress caused by wearing sensors attached to their bodies. Radar-based measurements have acquired physiological signals from animals including a rabbit and a rat~\cite{rabbit21}, a horse~\cite{horse23}, and a cow~\cite{cow19}. In this study, we have selected chimpanzees as the target animal because they are known to be the species most closely related to humans with the aim of developing an effective method for noncontact measurement of their heart inter-beat intervals (IBIs). 
Although camera-based measurements of chimpanzee cardiac signals~\cite{Wang2023} and radar-based measurement of chimpanzee's average heart rate~\cite{Matsumoto_Jxiv_2023} have been reported, this is the first report of radar-based accurate measurements of the time-dependent heart inter-beat intervals of a chimpanzee.

One of the greatest challenges with regard to radar-based heartbeat measurement is the effect of respiration because the body displacement caused by a heartbeat is much smaller than that caused by respiration \cite{heartdisp31,respdisp32}.
Several studies have therefore aimed to eliminate the respiratory component from the radar signals~\cite{MachineSpec1,PhaseVMD7,nophasezerocross15}.
Yamamoto and Otsuki used a band-pass filter with a machine learning algorithm to extract the heartbeat component of radar echo signals \cite{MachineSpec1}. Wang et al. used variational mode decomposition to separate the heartbeat and respiration components of radar echo signals \cite{PhaseVMD7}. Petrovi\'{c} et al. used a band-pass filter bank that was applied to the radar echo signal without arctangent demodulation \cite{nophasezerocross15}. These methods largely rely on the fundamental frequency component of the heartbeat component, which is often masked by the higher-order harmonics of the respiratory component, thus reducing the accuracy of heart IBI estimation.

In this study, we propose a method to determine the optimum cutoff frequency for a high-pass filter for use with the topology method \cite{topology27}, which is known as a method for radar-based estimation of IBI using waveform features. In previous studies, Wu et al. \cite{CNN-topology18} used the topology method with a band-pass filter with cutoff frequencies that were selected empirically. Through numerical examples performed using measured radar data, we clarify that the cutoff frequency should be selected to reject the respiratory component of the signal while maintaining the higher-order harmonics of the heartbeat component. After confirming the effectiveness of the proposed method by performing measurements on humans, we apply the same method to two chimpanzees under anesthesia.

\section{System Model and Topology Method}
\subsection{Radar System and Preprocessing}
We use a millimeter-wave radar system with a linear multiple-input multiple-output (MIMO) antenna array containing $N_\mathrm{T}$ transmitting elements and $N_\mathrm{R}$ receiving elements that can be approximated using $N (= N_\mathrm{T} N_\mathrm{R})$ virtual elements with spacing of one half of the wavelength $\lambda/2$.
Let $s_n(t,r)$ be the signal received when using the $n$-th virtual element, where $t$ is the slow time and $r=c\tau/2$ is the range that is proportional to both the fast time $\tau$ and the speed of light $c$.
Note that when a frequency-modulated continuous-wave radar system is used, the intermediate-frequency signals are converted into the fast-time domain using a Fourier transform.
Using Taylor weights $\alpha_n$ $(n=1,2,\cdots,N)$ with a beamformer weight vector $\vct{w}(\theta)=[w_1(\theta),w_2(\theta),\cdots,w_N(\theta)]^\mathrm{T}$, where $w_n(\theta)=\alpha_n\ee^{\jj \pi n \sin\theta}$, the signal vector $\vct{s}(t,r)=[s_1(t,r),s_2(t,r),\cdots,s_N(t,r)]^\TT$ can be converted into a radar image $I'(t,r,\theta)=\vct{w}^\HH(\theta) \vct{s}(t,r)$, where the superscripted $\TT$ and $\HH$ denote matrix/vector transposes and conjugate transpose, respectively.

Next, we suppress any static clutter components to obtain a clutter-free radar image $I(t,r,\theta)=I'(t,r,\theta)-(1/T)\int_0^T I'(t,r,\theta)\dd t$, where $T$ is the time for which the target position remains stationary.
The position of the target person/animal is then estimated to be $(r_0, \theta_0)=\arg\max_{(r,\theta)}\int_0^{T_0}|I(t,r,\theta)|^2\dd t$.
The displacement waveform is estimated to be $d'(t)=(\lambda/4\pi)\mathrm{unwrap}(\angle I(t,r_0,\theta_0))$, where $\angle$ denotes the phase of a complex number and unwrap denotes a phase unwrapping operator.
Finally, the trend component is estimated by performing smoothing using a Gaussian filter $g(t) = (1/\sqrt{2\pi}\sigma)\ee^{-t^2/2\sigma^2}$, and the physiological signal component $d(t)$ is estimated to be $d(t) = d'(t) - g*d'(t)$, where $*$ denotes a convolution operator.
Specifically, the parameters are set to be $N_\mathrm{T}=3$, $N_\mathrm{R}=4$, $N=12$, $\lambda=3.8$ mm, $T=T_0=120$ s for human measurements, $T=T_0=60$ s for the chimpanzee measurements, and $\sigma=1.0$ s in this study.

\subsection{Proposed Filter Parameter for Topology Method}
The topology method~\cite{topology27} is known to be an accurate method for radar-based heartbeat measurements. The topology method extracts six types of feature point (peak, trough, and four types of inflection points) from the body displacement waveform $d(t)$ and then introduces a new index called topology similarity, which acts as a measure of the repeatability of the feature sequences. Using the topology correlation, unreliable estimates can then be excluded to achieve accurate heart IBI estimation.

In the topology method, the IBI $\tau(t)$ is estimated from a pair of corresponding feature points (i.e., the $m$-th and $n$-th feature points) to be $\tau((t_m+t_n)/2) = t_n-t_m$ $(t_n > t_m)$ if the local correlation coefficient $C(m,n)$ and the topological similarity $M(m,n)$ satisfy $C(m,n)\geq C_0$ and $M(m,n)\geq M_0$, respectively, where $C_0$ and $M_0$ are thresholds.
If these thresholds $C_0$ and $M_0$ are set to have low values, many of the estimates of $\tau(t)$ then remain intact (i.e., $C(m,n)\geq C_0$ and $M(m,n)\geq M_0$), resulting in low estimation accuracy; if the thresholds are set to have high values, then many of the estimates of $\tau(t)$ are excluded (i.e., $C(m,n)<C_0$ or $M(m,n)<M_0$), which results in high accuracy for a limited number of estimates with many missing data ($\tau(t)$ that are not defined for numerous discrete time samples $t_i$); this indicates that there is a trade-off relationship between the estimation accuracy and the number of missing data.

One of the main challenges in taking radar-based heartbeat measurements is the interference between the respiration and heartbeat components. In particular, the heartbeat's fundamental frequency component (1.0--1.7 Hz for humans and 1.5--2.2 Hz for chimpanzees) is known to be affected by respiratory harmonics and intermodulation, thus making it difficult to separate or discriminate the measured components~\cite{higherharmonic26}.
To overcome this issue, we should use both the fundamental frequency of the heartbeat and its higher harmonics \cite{nophasezerocross15, higherharmonic26, explainresphamonic17, Higherharmonic16}. It was reported previously that the higher harmonics of the heartbeat component are much stronger than the higher (e.g., 8--10-th orders) harmonics of the respiration signal \cite{explainresphamonic17,nophasezerocross15}.
For the reasons described above, the higher harmonics of the heartbeat component can be identified with relatively little effort when compared with the fundamental frequency component, which is masked by the respiratory components.
In addition, it was reported that use of the second harmonic frequency of the heartbeat serves as an effective measure for suppression of the respiratory interference in radar data~\cite{higherharmonic26}.

In the proposed method, which is based on an estimate of the power spectral density $D(f)
  = |\mathcal{F}[d(t)]|^2$ of the estimated displacement $d(t)$, the second harmonic frequency $f_\mathrm{H2}=2f_\mathrm{H1}$ of the heartbeat is identified using the typical heart rate range for each species. We then detect the frequency $f_\mathrm{c}$ that corresponds to the trough of $D(f)$ and satisfies $(\dd/\dd f)D(f)|_{f=f_\mathrm{c}}=0$, $(\dd^2/\dd f^2)D(f)|_{f=f_\mathrm{c}}>0$, and $f_\mathrm{c} < f_\mathrm{H2}$, where $D(f)$ is smoothed appropriately. If multiple candidates for $f_\mathrm{c}$ are available, we select the $f_\mathrm{c}$ that minimizes $|f_\mathrm{H2}-f_\mathrm{c}|$. We then apply a high-pass filter with a cutoff frequency $f_\mathrm{c}$ to the displacement waveform $d(t)$; the resulting waveform is used as the input for the topology method.
In the next section, we evaluate the performance of the proposed method using the results of measurements involving both human and chimpanzee subjects.

\section{Evaluation of Estimation Accuracy of Proposed Method}
\subsection{Accuracy Evaluation with Human Participants}
\begin{figure}[tb]
  \centering
  \includegraphics[width =0.5\linewidth,pagebox=cropbox,clip]{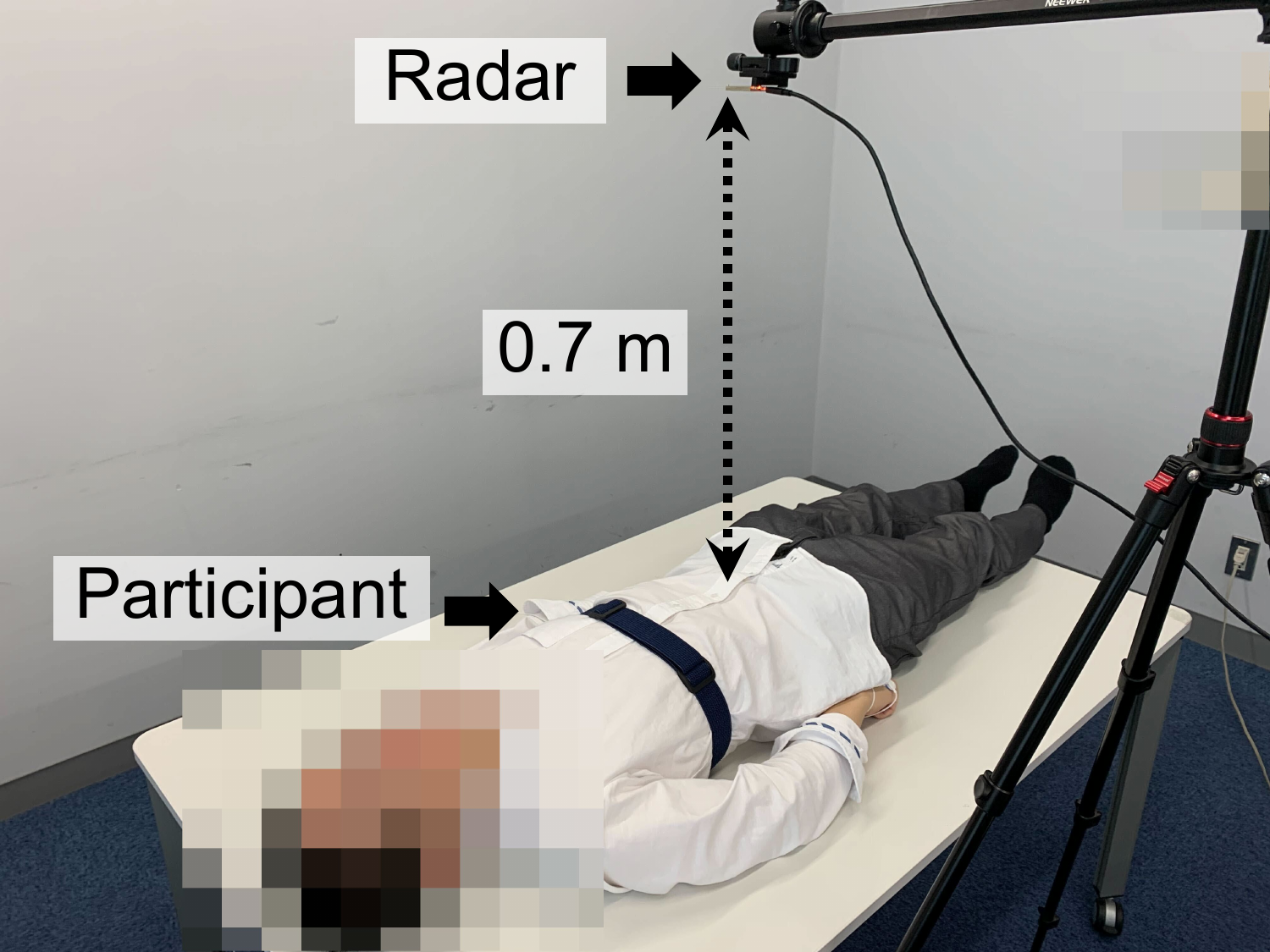}
  \caption{Photograph of the measurement setup with a human participant.}
  \label{fig:radar_measurement_setup_human}
\end{figure}

To evaluate the effectiveness of the proposed method, preliminary experiments were conducted with two human participants. The participants were placed in a supine position and breathed naturally for 2 min with the electrodes of an electrocardiogram (ECG) sensor attached to their chest (see Fig. \ref{fig:radar_measurement_setup_human}). Note that the ECG sensor was only used here to evaluate the accuracy of the radar-based measurements.

We used a frequency-modulated continuous-wave radar system with a center frequency of 79 GHz, a center wavelength $\lambda = 3.8$ mm, a bandwidth of 3.5 Hz, and a slow-time sampling frequency of 100 Hz. 
The radar has a MIMO antenna array that consists of three transmitting and four receiving elements with intervals of 7.6 mm ($2\lambda$) between the transmitting elements and 1.9 mm ($\lambda/2$) between the receiving elements; this setup can be approximated using a 12-element virtual linear array. 
We placed the radar at a distance of 0.7 m from each participant and ensured that the participant's body was oriented in the direction of the array baseline.

\begin{figure}[tb]
  \centering
  \includegraphics[width =0.65\linewidth,pagebox=cropbox,clip]{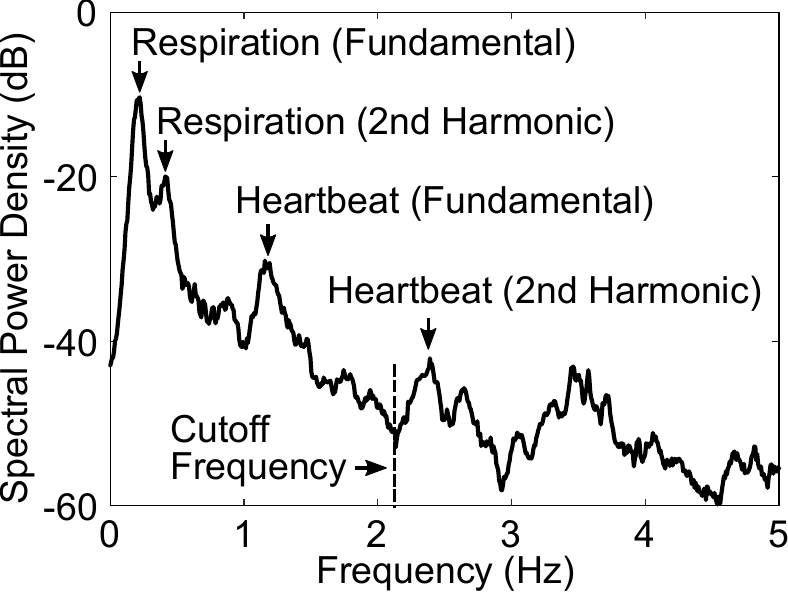}
  \caption{Estimated power spectral density $D(f)$ characteristic of the displacement $d(t)$ for participant A.}
  \label{fig:wang_2nd_fft_periodogram}
\end{figure}

\begin{figure}[tb]
  \begin{minipage}[tb]{\linewidth}
    \centering
    \includegraphics[width = 0.8\linewidth,pagebox=cropbox,clip]{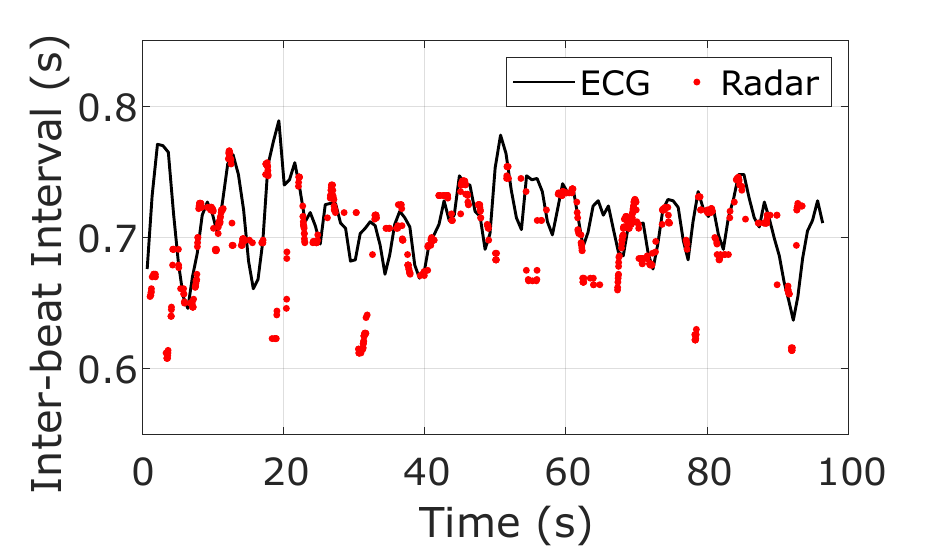}
    \subcaption{IBI estimated using conventional filter (participant A).}
    \label{fig:wang_1st_conventional18_topology_result}
  \end{minipage}
  \begin{minipage}[tb]{\linewidth}
    \centering
    \includegraphics[width = 0.8\linewidth,pagebox=cropbox,clip]{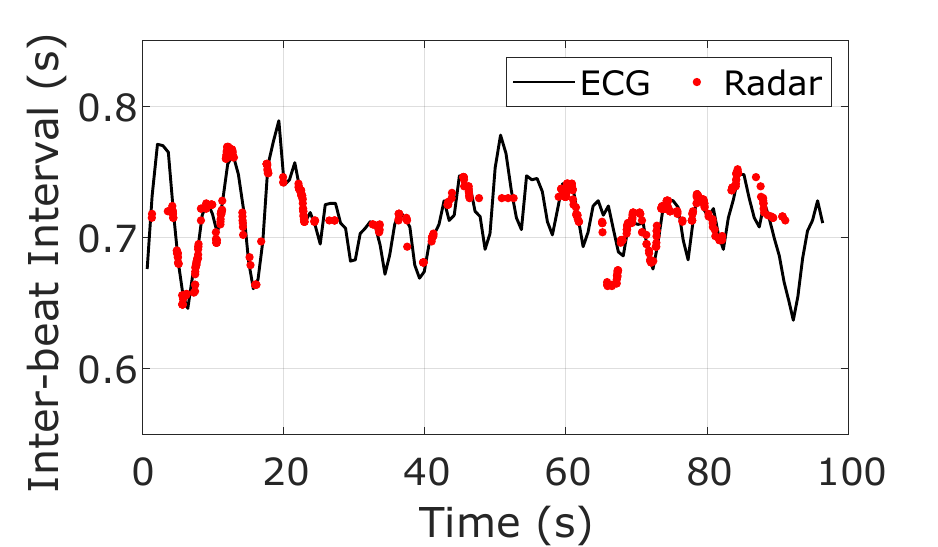}
    \subcaption{IBI estimated using proposed filter (participant A).}
    \label{fig:wang_1st_topology_result_prop}
  \end{minipage}
  \caption{IBI characteristics estimated using radar and the topology method with (a) conventional filter \cite{CNN-topology18} and (b) proposed filter, where the black lines are reference IBIs obtained from the ECG.}
  \label{fig:human_topology_result_comparison}
\end{figure}

Fig.~\ref{fig:wang_2nd_fft_periodogram} shows the power spectral density $D(f)$ of the displacement $d(t)$ for participant A, which also shows a dashed line that indicates the proposed cutoff frequency $f_\mathrm{c}$ for a high-pass filter. The figure also shows the fundamental and higher harmonics for each component.
In the figure, the fundamental and second harmonic frequencies of the respiration and heartbeat components are shown clearly. However, in many cases, the fundamental frequency component of the heartbeat is often masked by the higher harmonics of the respiratory component.
Fig.~\ref{fig:human_topology_result_comparison} shows the IBI that was estimated using the topology method with a band-pass filter that had the same cutoff frequencies that were used in \cite{CNN-topology18}, and corresponding results obtained with a high-pass filter with the proposed cutoff frequency. The figure shows that the proposed method can achieve accurate estimation, particularly for $t \leq 20$ s and $60$ s $\leq t \leq 80$ s.

We use the root-mean-square (RMS) error as a performance index for both the conventional and proposed methods.
For the IBI $\tau(t)$ estimated using radar and the corresponding IBI $\tau'(t)$ obtained from the reference ECG, the RMS error is defined as $\sqrt{(1/T)\int_0^T |\tau(t)-\tau'(t)|^2\dd t}$.
As conventional methods for comparison, we used the topology method with two types of band-pass filter: (1) with cutoff frequencies covering only the fundamental heartbeat frequency, and (2) with the same cutoff frequencies as in~\cite{CNN-topology18}.
Table~\ref{tbl:comparison_human_topology} shows the RMS errors for the two conventional methods and for the proposed method for two participants, A and B. The proposed method demonstrates higher accuracy than the conventional methods. These results illustrate the effectiveness of the proposed method for accurate heartbeat measurements using radar.

\begin{table}[tb]
  \centering
  \caption{RMS error in IBI estimation for human participants.}\vspace{-5px}
  \scriptsize
  \begin{tabular}{c|ccc}
    \toprule
    \multirow{2}{*}{Participant} & \multicolumn{3}{c}{RMS error (ms)}                                             \\\cline{2-4}
                                 & Fundamental freq.                  & \cite{CNN-topology18} & \textbf{Proposed} \\
    \midrule
    A (1st trial)                & 55.10                              & 39.62                 & \textbf{11.03}    \\
    A (2nd trial)                & 42.09                              & 6.08                  & \textbf{4.43}     \\
    B (1st trial)                & 122.3                              & 75.22                 & \textbf{50.91}    \\
    B (2nd trial)                & 82.14                              & 48.83                 & \textbf{26.03}    \\
    \bottomrule
  \end{tabular}
  \normalsize
  \label{tbl:comparison_human_topology}
\end{table}

\begin{figure}[tb]
  \centering
  \includegraphics[width = 0.65\linewidth,pagebox=cropbox,clip]{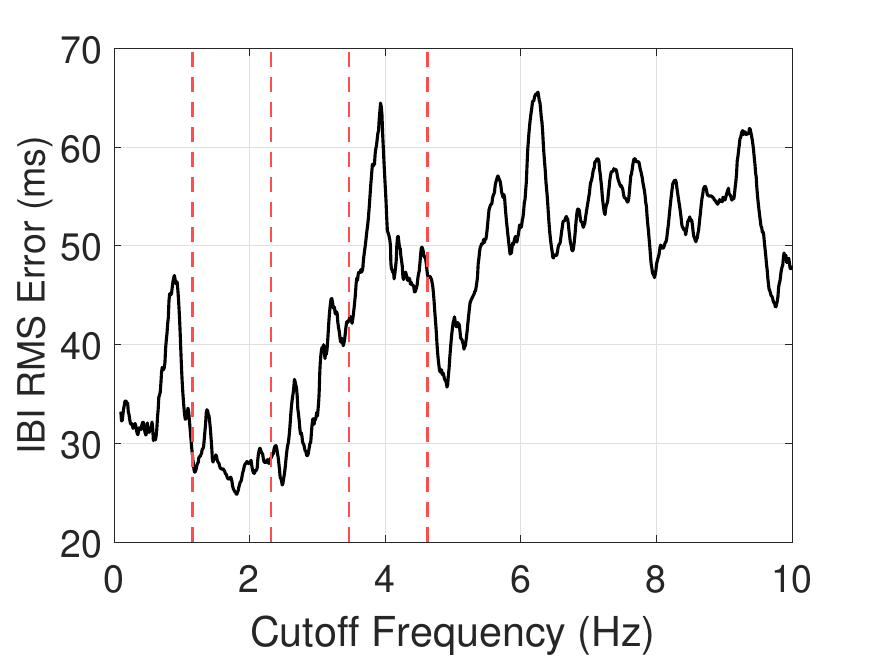}
  \caption{RMS error (black line) in IBI estimation for the cutoff frequency $f_\mathrm{c}$ of the high-pass filter. Red dashed lines indicate the heartbeat's fundamental frequency and its harmonics.}
  \label{fig:RMSE-cutoff_freq}
\end{figure}

Fig.~\ref{fig:RMSE-cutoff_freq} shows the RMS error values versus the cutoff frequency $f_\mathrm{c}$ of the high-pass filter for participant A, where the red dashed lines indicate the higher harmonics of the heartbeat. The RMS error is shown to decrease when $f_\mathrm{c}$ is located slightly lower than each of the higher harmonics of the heartbeat, resulting in improved performance. These results validate our proposed method.

\subsection{Accuracy Evaluation with Chimpanzee Subjects}
Next, we performed radar measurements of two chimpanzee subjects (C and D) during the annual health checkups of these chimpanzees; the subjects were anesthetized before the radar measurements. Subjects C and D are adult male and female chimpanzees, respectively. 

We used a radar system that was almost the same as that used for the human participants in the previous subsection; the slow-time sampling frequency was 145.56 Hz in this case. The radar system was placed approximately 0.7 m away from the chimpanzee, immediately above the front chest wall, and the ECG electrodes were attached to one arm and one leg. 

\begin{figure}[tb]
  \centering
  \includegraphics[width = 0.8\linewidth,pagebox=cropbox,clip]{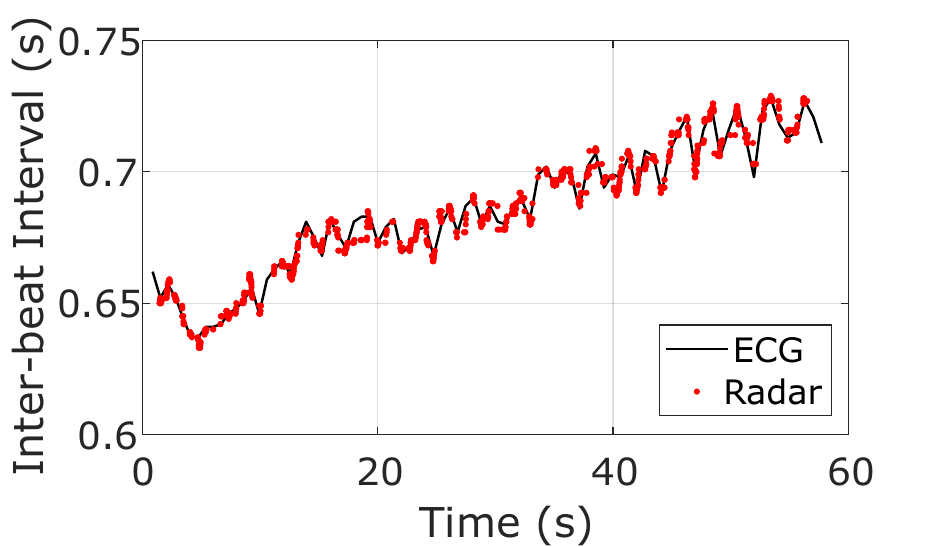}
  \caption{IBIs estimated using radar and the topology method with the proposed filter (red dots) and the reference IBI results obtained from the ECG (black line) for chimpanzee C.}
  \label{fig:chimpanzees_result}
\end{figure}

As in the preliminary human experiments, the proposed method was applied to the radar measurement data from the chimpanzees to obtain their IBIs.
Fig.~\ref{fig:chimpanzees_result} shows the IBI estimation results for chimpanzee C, which shows good agreement between the radar and ECG results. 
The RMS errors of IBI estimation were 2.6 ms and 20.5 ms for chimpanzee subjects C and D, respectively, from which we note that the accuracy for subject D was lower than that for subject C; we noticed large limb movements that occurred every 3--5 s, possibly because of the weak effect of the anaesthesia in this case. It will be important to develop a method to suppress the undesired components caused by such body movements, and this will form part of our future research.

\begin{figure}[tb]
  \centering
  \includegraphics[width = 0.4\linewidth,pagebox=cropbox,clip]{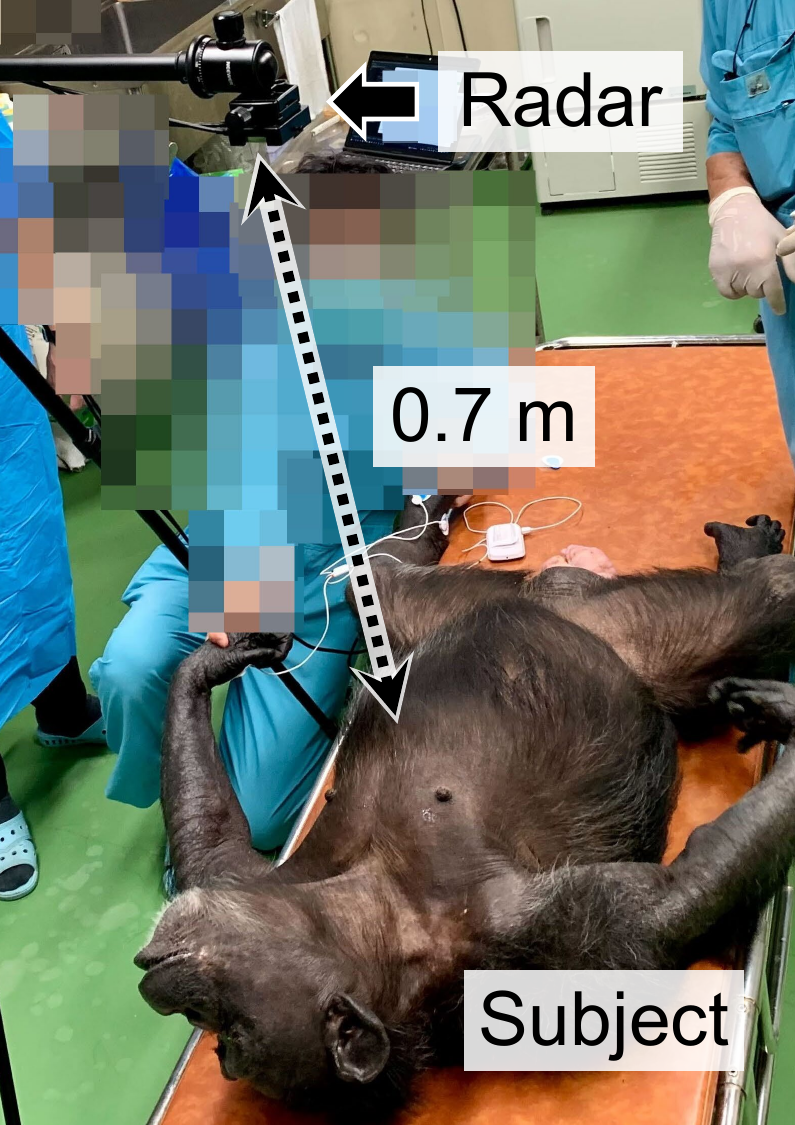}
  \caption{Photograph of an experiment involving radar measurement of a chimpanzee subject.}
  \label{fig:chimpanzee_experiment}
\end{figure}

\section{Conclusion}
In this study, we have proposed a method to set the cutoff frequency of a high-pass filter used for the topology method and then evaluated its effectiveness in noncontact measurement of the heart IBI of both human participants and chimpanzee subjects. The accuracy of IBI estimation was evaluated quantitatively and the results demonstrated the importance of appropriate selection of the cutoff frequency for suppression of the respiratory components while maintaining the required heartbeat components. Unlike conventional studies that use empirically set filter parameters, this approach introduced a systematic method for filter parameter setting based on the power spectral density of the displacement waveform. We also noted that the accuracy can deteriorate in the presence of body movements; this problem will be addressed in future studies using additional data taken from human participants and chimpanzee subjects.

\section*{Ethics Declarations}
The experimental protocol involving animals was approved by the Animal Experimentation Committee of the Wildlife Research Center, Kyoto University (WRC-2022-KS002A). The experimental protocol involving human participants was approved by the Ethics Committee of the Graduate School of Engineering, Kyoto University (permit no.~202214). Informed consent was obtained from all human participants in the study.

\section*{Acknowledgment}
\addcontentsline{toc}{section}{Acknowledgment}
This work was supported in part by the SECOM Science and Technology Foundation, by JST under Grant JPMJMI22J2, and by JSPS KAKENHI under Grants 19H02155, 20K20875, 21H03427, 23H01420, and 23H03881, Grant for Collaborative Research in Humanities and Social Sciences by Suntory Foundation, and Grant for Agri-transformation, Shinshu University. The authors thank Dr. Toshifumi Udono, Dr. Etsuko Nogami, Mr. Shunsuke Iwata, Mr. Haruto Kobayashi, and Mr. Yu Oshima of Kyoto University for their help with the measurements, and we also thank Dr. Hirofumi Taki and Dr. Shigeaki Okumura of MaRI Co., Ltd. for their technical advice. 

\end{document}